\documentclass{amsart}

\usepackage{dcolumn}

\usepackage{amsmath}

\DeclareMathOperator{\Spin}{Spin}
\DeclareMathOperator{\SU}{SU}

\usepackage[ps,matrix,arrow,poly,graph,curve,line]{xy}
\UsePSspecials{dvips}

\labelmargin-{1.2pt}  

\newcommand{\SL}{{\rm SL}} 

\newcommand{\SO}{{\rm SO}}

\newcommand{\R}{{\mathbb R}}  
\newcommand{\C}{{\mathbb C}}  
\newcommand{\Z}{{\mathbb Z}}  
\newcommand{\D}{{\mathcal D}}
  
\renewcommand{\H} {{H}}  
\newcommand{\maps}{\colon} 
\newcommand{\tensor}{\otimes}

\newtheorem{lemma}{Lemma}  
\newtheorem{corollary}{Corollary}
\newtheorem{theorem}{Theorem}

\newcommand{\cxymatrix}[1]{\vcenter{\xymatrix{#1}}}
  
\newcommand{\fifteenj}{%
\def\lab{\ifcase\xypolynode\or k_{1} \or j_{1} \or k_{2} \or j_{2} \or k_{3} \or j_{3} \or k_{4} \or j_{4} \or k_{5} \or j_{5} \fi}
\begin{xy} 
\xygraph{!{<3.2pc,0pc>:}
  !P10"A"{~><{@{{}{-}*{\bullet}}} ~>>{_\lab}}
  "A2" -@-_{j_6} "A5"
  "A6" -@-_{j_8} "A9"
  "A10" -@-_{j_{10}} "A3"
  "A4" -@-_{j_7} "A7"
  "A8" -@-_{j_9} "A1"
}
\end{xy}
}

\newcommand{\TenJ}{%
\begin{xy} 
\xygraph{!{<4pc,0pc>:}
  !P5"A"{~><{@{{}{-}*{\bullet}}} ~>>{_{j_{\xypolynode}}}}
  "A1" -@-_{j_6} "A3"
  "A2" -@-_{j_7} "A4"
  "A3" -@-_{j_8} "A5"
  "A4" -@-_{j_9} "A1"
  "A5" -@-_{j_{10}} "A2"
}
\end{xy}
}

\newcommand{\ModifiedTenJ}{%
\begin{xy} 
\xygraph{!{<4pc,0pc>:}
  !P5"A"{~><{@{{}{-}*{\bullet}}} ~>>{_{j_{\xypolynode}}} ~<>{?(0.74);?(1.2) **\dir{..}} }
  "A1" -@-_{j_6} "A3"
  "A2" -@-_{j_7} "A4"
  "A3" -@-_{j_8} "A5"
  "A4" -@-_{j_9} "A1"
  "A5" -@-_{j_{10}} "A2"
}
\end{xy}
}

\newcommand{\TenJJ}{%
\def\lab{\ifcase\xypolynode\or 0,1 \or 1,2 \or 2,3 \or 3,4 \or 4,0 \fi}
\begin{xy} 
\xygraph{!{<4pc,0pc>:}
  !P5"A"{~><{@{{}{-}*{\bullet}}} ~>>{_{j_{\lab}}}}
  "A1" -@-_{j_{0,2}} "A3"
  "A2" -@-_{j_{1,3}} "A4"
  "A3" -@-_{j_{2,4}} "A5"
  "A4" -@-_{j_{3,0}} "A1"
  "A5" -@-_{j_{4,1}} "A2"
}
\end{xy}
}

\newcommand{\FourX}[4]  
{  
\cxymatrix{\ar@{-}[dr]^{#1} & & \ar@{-}[dl]_{#2} \\  
 & *{\bullet} & \\  
\ar@{-}[ur]_{#3} & & \ar@{-}[ul]^{#4}\\}  
}  
 
\newcommand{\FourXh}[4]  
{  
\cxymatrix{\ar@{-}[dr]^{#1} & & \ar@{-}[dl]_{#2} \\  
 & *{\bullet} \ar@{-}[l] \ar@{-}[r] & \\  
\ar@{-}[ur]_{#3} & & \ar@{-}[ul]^{#4}\\}  
}  
 
\newcommand{\FourXhdashed}[4]  
{  
\cxymatrix{\ar@{-}[dr]^{#1} & & \ar@{-}[dl]_{#2} \\  
 & *{\bullet} \ar@{--}[]-<12pt,0pt> \ar@{--}[]+<12pt,0pt> & \\  
\ar@{-}[ur]_{#3} & & \ar@{-}[ul]^{#4}\\}  
}  
 
\newcommand{\FourXhdotted}[4]  
{  
\cxymatrix{\ar@{-}[dr]^{#1} & & \ar@{-}[dl]_{#2} \\  
 & *{\bullet} \ar@{{}*{\cdot }{}}[]-<12pt,0pt> \ar@{{}*{\cdot }{}}[]+<12pt,0pt> & \\  
\ar@{-}[ur]_{#3} & & \ar@{-}[ul]^{#4}\\}  
}  
 
\newcommand{\FourXv}[4]  
{  
\cxymatrix{\ar@{-}[dr]^{#1} & & \ar@{-}[dl]_{#2} \\  
 & *{\bullet} \ar@{-}[d] \ar@{-}[u] & \\  
\ar@{-}[ur]_{#3} & & \ar@{-}[ul]^{#4}\\}  
}  
 
\newcommand{\FourXvdashed}[4]  
{  
\cxymatrix{\ar@{-}[dr]^{#1} & & \ar@{-}[dl]_{#2} \\  
 & *{\bullet} \ar@{--}[]-<0pt,12pt> \ar@{--}[]+<0pt,12pt> & \\  
\ar@{-}[ur]_{#3} & & \ar@{-}[ul]^{#4}\\}  
}  
 
\newcommand{\FourXvdotted}[4]  
{  
\cxymatrix{\ar@{-}[dr]^{#1} & & \ar@{-}[dl]_{#2} \\  
 & *{\bullet} \ar@{{}*=<0pt,3pt>{\cdot}{}}[]-<0pt,12pt> \ar@{{}*=<0pt,3pt>{\cdot}{}}[]+<0pt,12pt> & \\  
\ar@{-}[ur]_{#3} & & \ar@{-}[ul]^{#4}\\}  
}  
 
\newcommand{\DoubleY}[5] 
{  
\cxymatrix{\ar@{-}[dr]^{#1} & & \ar@{-}[dl]_{#2} \\  
 & *{\bullet} \ar@{-}[d]^{#5} \\
 & *{\bullet} \\
\ar@{-}[ur]_{#3} & & \ar@{-}[ul]^{#4}}  
}  
 
\newcommand{\DoubleYhor}[5] 
{  
\cxymatrix{\ar@{-}[dr]^{#1} & & & \ar@{-}[dl]_{#2} \\  
 & *{\bullet} \ar@{-}[r]^{#5}& *{\bullet} &\\  
\ar@{-}[ur]_{#3} & & & \ar@{-}[ul]^{#4}}  
}  
 
\newcommand{\monogon}[1]  
{  
\cxymatrix{ *{\bullet}  
\ar@  
{-}  
@(ul,dl)  
[]  
_{#1}  
\\}  
}  
 
\newcommand{\bigon}[2]  
{  
\cxymatrix{ *{\bullet} 
\ar@{-} 
@/^1pc/ 
[r] 
^{#1} 
\ar@{-} 
@/_1pc/ 
[r] 
_{#2} 
& 
*{\bullet} 
} 
} 

\newcommand{\thetagraph}[3]  
{ 
\cxymatrix{ *{\bullet} 
\ar@{-} 
@/^1.5pc/ 
[r] 
^{#1} 
\ar@{-} 
@/_1.5pc/ 
[r] 
_{#2} 
\ar@{-} 
[r]^{#3} 
& 
*{\bullet} 
\\} 
} 

\newcommand{\unigon}[1]  
{ 
\cxymatrix{ *{\bullet} 
\ar@{-} 
[r]^{#1}  
& 
*{\bullet} 
\\} 
}

\newcommand{\fourtheta}[4]  
{ 
\cxymatrix{ *{\bullet} 
\ar@{-} 
@/^1.5pc/ 
[r] 
^{#1} 
\ar@{-} 
@/_1.5pc/ 
[r] 
_{#2} 
\ar@{-} 
@/^/ 
[r]^{#3} 
\ar@{-} 
@/_/ 
[r]_{#4} 
& 
*{\bullet} 
\\} 
} 

\begin{document} 

\title{Positivity of Spin Foam Amplitudes }   
\author{John C.\ Baez}
\address{John C.\ Baez \\
Department of Mathematics\\
University of California\\
Riverside, California 92521}
\email{baez@math.ucr.edu}
\author{J.\ Daniel Christensen}
\address{J.\ Daniel Christensen \\
Department of Mathematics\\
University of Western Ontario\\
London, ON N6A 5B7 Canada}
\email{jdc@uwo.ca}

\date{February 21, 2002}

\begin{abstract}     
The amplitude for a spin foam in the Barrett--Crane model of Riemannian
quantum gravity is given as a product over its vertices, edges and
faces, with one factor of the Riemannian $10j$ symbols appearing for
each vertex, and simpler factors for the edges and faces.  We prove that
these amplitudes are always nonnegative for closed spin foams.  As a
corollary, all open spin foams going between a fixed pair of spin networks 
have real amplitudes of the same sign.  This
means one can use the Metropolis algorithm to compute expectation values
of observables in the Riemannian Barrett--Crane model, as in statistical
mechanics, even though this theory is based on a real-time ($e^{iS}$)
rather than imaginary-time ($e^{-S}$) path integral.  Our proof uses
the fact that when the Riemannian $10j$ symbols are nonzero, their sign
is positive or negative depending on whether the sum of the ten spins is
an integer or half-integer.  For the product of $10j$ symbols appearing
in the amplitude for a closed spin foam, these signs cancel.  We
conclude with some numerical evidence suggesting that the Lorentzian
$10j$ symbols are always nonnegative, which would imply similar results
for the Lorentzian Barrett--Crane model.
\end{abstract}

\maketitle

\section{Introduction}

Physicists have long found it
difficult to deal with the oscillatory behavior
of the integrand in real-time Feynman path integrals of the form
\[        \int f(x) \, e^{iS(x)} \, \D x. \]
For this reason, we have become accustomed to using Wick rotation to
switch to imaginary time, thus replacing these integrals by those of
the form
\[        \int f(x)\, e^{-S(x)} \, \D x \]
where the positivity of the exponential allows one to use concepts from
probability theory.  This is helpful not only in analytical work but
also in numerical computations.  For example, computations in lattice
field theory rely heavily on techniques such as the Metropolis algorithm.
They give powerful tools for computing the latter sort of integral, 
but are typically useless for those of the former sort.

Unfortunately, the strategy of Wick rotation becomes problematic in
the context of quantum gravity.  The basic difficulty is that in the
absence of a favored time coordinate, the replacement $t \mapsto it$
makes no sense unless one can prove that the results obtained thereby
are independent of the choice of the $t$ coordinate.  A great deal of
work on `Euclidean quantum gravity' is of dubious relevance to
real-world physics because of inattention to this issue.  

Luckily, in the absence of topology change (i.e.\ for spacetimes of
the topology $\R \times S$) there exist theorems justifying Wick
rotation for certain diffeomorphism-invariant field theories
\cite{AMMT}.  Taking the prohibition against topology change seriously
has led to models of quantum gravity in which Wick rotation is indeed
justified, and these appear to be better-behaved than
those previously considered in Euclidean quantum gravity
\cite{AJL1,AJL2}.

Nonetheless, there are still reasons to be interested in real-time
Feynman path integrals in quantum gravity.  After all, the prohibition
against topology change might be a limitation of the Wick rotation
method rather than an actual law of physics.  Moreover, the original
Feynman integrals have a certain conceptual priority over their
Wick-rotated counterparts: quantum theory is ultimately about
amplitudes, not probabilities.

Yet another reason is that there are some very interesting models of
quantum gravity based on real-time path integrals: namely, spin foam
models \cite{Baez,Baez2,BC,BC2,Oriti,PR,RR}.  In a spin foam model of
quantum gravity, the real-time path integral is rewritten as a sum 
over `spin foams': 2-dimensional cell complexes with faces labelled by
group representations and edges labelled by intertwiners.  This sum has
been rigorously shown to converge in some of these models \cite{CPR,P},
at least for a fixed triangulation of spacetime.  The next step is to
determine whether these models have good behavior at large distance
scales: that is, whether they reduce to general relativity in a suitable
limit.  For this, we need to compute expectation values of observables,
e.g.\ by numerical methods.  

Here we show that for a specific model of this sort --- the Riemannian
Barrett--Crane model --- one can do such computations using the
Metropolis algorithm, because the amplitudes for closed spin foams are
{\it nonnegative}.  This model is not physically realistic: it is a
version of `Riemannian quantum gravity', that is, the quantum field
theory based on a real-time path integral involving the
Einstein--Hilbert action for Riemannian metrics.  Since the Cauchy
problem is ill-posed in classical Riemannian general relativity, we
should expect to run into  problems with the quantum theory at some
point.  Nonetheless, the Riemannian Barrett--Crane model has so far
proved to be a useful guide to  work on the physically more important
but mathematically more challenging Lorentzian Barrett--Crane model. 
In particular, we conjecture that the nonnegativity of spin foam
amplitudes holds for the Lorentzian model as well as the Riemannian
one.

In Section 2 we prove the nonnegativity of spin foam amplitudes in two
versions of the Riemannian Barrett--Crane model: the version due to De
Pietri, Freidel, Krasnov and Rovelli \cite{DFKR}, and the version with
modified edge and face amplitudes due to Perez and Rovelli \cite{PR}. In
the Perez--Rovelli version the partition function has been shown to
converge \cite{P}; in the DFKR version the partition function appears to
diverge \cite{BaezChristensen}.  However, in both versions the spin foam
vertex amplitudes are given by the Riemannian $10j$ symbols, and the key
fact we need to show is that the product of these $10j$ symbols over all
vertices of a closed spin foam is nonnegative.  We prove this in two
steps.  In Lemma \ref{lemma4}, we show that when the Riemannian $10j$
symbol is nonzero, its sign is positive or negative depending on whether
the sum of the ten spins in question is an integer or half-integer.  In
Lemma \ref{lemma5}, we show that all the resulting minus signs cancel
when we take the product of $10j$ symbols over all vertices of a closed
spin foam.  This immediately gives the desired result.  As a corollary
of this result, we show that all open spin foams going between fixed
spin networks have real amplitudes of the same sign.  Finally, we
present some numerical evidence supporting our conjecture that the
Lorentzian $10j$ symbols are always nonnegative.

\section{The Riemannian Model}

In what follows we shall freely make use of standard graphical notation 
for $\SU(2)$ spin networks \cite{CFS,KL}.   We normalize trivalent
vertices so that
\[
\thetagraph{j_1}{j_2}{j_3} = 1
\]
whenever the spins $j_1,j_2,j_3$ form an admissible triple: that is,
satisfy the triangle inequality and sum to an integer.  (When the spins
do not form an admissible triple, we define the trivalent vertex to be
zero.)  As this normalization sometimes requires dividing by the square
root of a negative number, it introduces a potential sign ambiguity.
Luckily, in our calculations trivalent vertices always come in matching
pairs, so these signs cancel, leaving an unambiguous result.  This is
important, because signs are of the essence in all to come.

We shall also use the standard notation for simple spin networks
\cite{BC,Yetter}.  In such a spin network, labelling an edge by the spin
$j$ really means that it is labelled by irreducible representation $j
\tensor j$ of the group $\Spin(4) = \SU(2) \times \SU(2)$.  Such
representations are called `simple', and they all descend to
representations of $\SO(4)$.  Unlike $\SU(2)$ spin networks, which
change sign when one inserts a twist in the framing of any edge labelled
by a half-integer spin, simple spin networks are framing-independent.  
In a simple spin network, any unlabelled 4-valent vertex is really labelled 
by the Barrett--Crane intertwiner.  This intertwiner is defined in terms of 
$\SU(2)$ intertwiners by:
\[
\FourX{j_1}{j_2}{j_3}{j_4} = \quad \sum_j (-1)^{2j} (2j+1) \;
\DoubleY{j_1}{j_2}{j_3}{j_4}{j} \tensor
\DoubleY{j_1}{j_2}{j_3}{j_4}{j} 
\]
where we sum over spins $j$ such that the triples $j_1,j_2,j$ and
$j_3,j_4,j$ are both admissible.   Superficially, this definition
involves splitting the four edges incident to the vertex 
into a pair of `incoming' edges and a pair of `outgoing' ones.  
However, the Barrett--Crane intertwiner is actually independent
of this choice.  For example, we also have
\[
\FourX{j_1}{j_2}{j_3}{j_4} = \,\, \sum_j  (-1)^{2j} (2j+1)  \;
\DoubleYhor{j_1}{j_2}{j_3}{j_4}{j} \tensor
\DoubleYhor{j_1}{j_2}{j_3}{j_4}{j} 
\]
This fact is stated in Barrett and Crane's original paper
\cite{BC}; a formal proof was given by Yetter \cite{Yetter}.

It will be technically useful to modify the Barrett--Crane intertwiner by
removing the signs from the above formulas.  Our final results will
concern the original Barrett--Crane intertwiner, but our intermediate
calculations will use the intertwiner without signs.
In this modified intertwiner we use a dotted line to
indicate the splitting of incident edges into `incoming' and
`outgoing' pairs:
\[
\FourXhdotted{j_1}{j_2}{j_3}{j_4} = \quad \sum_j (2j+1) \;
\DoubleY{j_1}{j_2}{j_3}{j_4}{j} \tensor
\DoubleY{j_1}{j_2}{j_3}{j_4}{j} 
\]
In fact, this modified intertwiner only differs from the 
usual one by an overall sign.

\begin{lemma} \label{lemma1} 
\[
\FourXhdotted{j_1}{j_2}{j_3}{j_4} 
= \quad (-1)^{2(j_1 + j_2)} \FourX{j_1}{j_2}{j_3}{j_4}  
= \quad (-1)^{2(j_3 + j_4)} \FourX{j_1}{j_2}{j_3}{j_4}  
\]
It follows that
\[  \FourX{j_1}{j_2}{j_3}{j_4} = 0 \]
unless $j_1 + \cdots + j_4$ is an integer.
\end{lemma}

Proof --- Since the triples $j_1,j_2,j$ and 
$j_3,j_4,j$ are admissible in the sum 
\[
\FourX{j_1}{j_2}{j_3}{j_4} = \quad \sum_j (-1)^{2j} \, (2j+1)  \;
\DoubleY{j_1}{j_2}{j_3}{j_4}{j} \tensor
\DoubleY{j_1}{j_2}{j_3}{j_4}{j} 
\]
we have $(-1)^{2j} = (-1)^{2(j_1 + j_2)} = (-1)^{2(j_3 + j_4)}$, 
which yields the equations to be proved.
\qed
\vskip 1em

This means that the modified intertwiner changes at most by
a sign when we change the splitting.  For example, consider
\[
\FourXvdotted{j_1}{j_2}{j_3}{j_4} = \quad \sum_j (2j+1) \;
\DoubleYhor{j_1}{j_2}{j_3}{j_4}{j} \tensor
\DoubleYhor{j_1}{j_2}{j_3}{j_4}{j} 
\]
This satisfies:

\begin{lemma} \label{lemma2} 
\[ 
\FourXvdotted{j_1}{j_2}{j_3}{j_4} 
= \quad (-1)^{2(j_1 + j_4)} \FourXhdotted{j_1}{j_2}{j_3}{j_4} 
= \quad (-1)^{2(j_2 + j_3)} \FourXhdotted{j_1}{j_2}{j_3}{j_4} 
\]
\end{lemma}

Proof --- By Lemma \ref{lemma1} we have
\[
\FourXhdotted{j_1}{j_2}{j_3}{j_4} =  \quad
(-1)^{2(j_1 + j_2)} \FourX{j_1}{j_2}{j_3}{j_4} 
\]
Rotating the pictures clockwise and relabelling, we have
\[
\FourXvdotted{j_1}{j_2}{j_3}{j_4} = \quad
(-1)^{2(j_2 + j_4)} \FourX{j_1}{j_2}{j_3}{j_4} 
\]
Combining these equations, it follows that
\[
\FourXhdotted{j_1}{j_2}{j_3}{j_4} = \quad
(-1)^{2(j_1 + j_4)} \FourXvdotted{j_1}{j_2}{j_3}{j_4} 
\]
as desired.  A similar argument where we rotate counterclockwise
gives
\[
\FourXhdotted{j_1}{j_2}{j_3}{j_4} =  \quad
(-1)^{2(j_2 + j_3)} \FourXvdotted{j_1}{j_2}{j_3}{j_4} 
\]
\hskip 50 em \qed
\vskip 1em

The usual Riemannian $10j$ symbol \cite{BC} is the function of 10 spins 
obtained by evaluating the following closed $\SU(2) \times \SU(2)$ 
spin network:
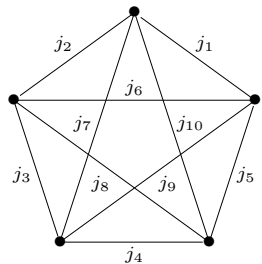
\[    \TenJ \]
where the vertices are labelled by Barrett--Crane intertwiners.
Mimicking this construction, we can also define a modified $10j$ 
symbol using the modified Barrett--Crane intertwiners:
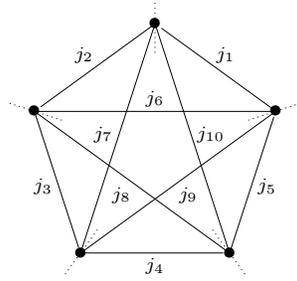
\[  \ModifiedTenJ  \]
Here the choice of splittings really matters: by Lemma \ref{lemma2}, if
we change the splitting at any vertex, the value of the spin network
might change sign.  The nice thing about the above choice is that it 
always gives a {\it nonnegative} result:

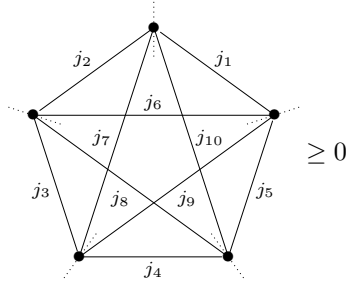
\begin{lemma} \label{lemma3} 
\[    \ModifiedTenJ  
\ge 0 \]
\end{lemma}

Proof --- Using the definition of the modified
Barrett--Crane intertwiner, the modified $10j$ symbol above
can be rewritten as the following weighted sum of $\SU(2)$ spin networks:
\[  \sum_{k_1,k_2,k_3,k_4,k_5} \, ({\prod_{i=1}^5 \, (2k_i+1)})\,
\begin{matrix}
{\fifteenj \, \fifteenj}
\end{matrix}
\]
Since the value of a disjoint union of spin networks is the product
of their values, it follows that
\begin{equation}
\ModifiedTenJ 
= \sum_{k_1,\dots,k_5}  ({\prod_{i=1}^5 \, (2k_i+1)})
\left(\fifteenj\right)^{2}
\label{sum.of.squares}
\end{equation}
To show that the modified $10j$ symbols are positive, it
thus suffices to show that this $\SU(2)$ spin network,
called a $15j$ symbol:
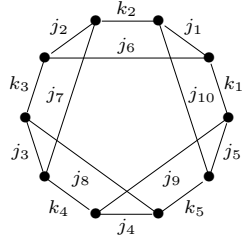
\[
\fifteenj
\]
evaluates to a real number.  

To see this, first note that the only place where complex numbers are
essential to the theory of trivalent $\SU(2)$ spin networks is in the
normalization of the trivalent vertex.  If we used the projection from
the tensor product $j_1 \tensor j_2$ to the summand $j_3$ as our
trivalent vertex whenever $j_1,j_2,j_3$ is an admissible triple, all
closed trivalent $\SU(2)$ spin networks would evaluate to real numbers,
since we could define them using the representation theory of
$\SL(2,\R)$ instead of $\SU(2)$.  However, this would not give
\[
\thetagraph{j_1}{j_2}{j_3} = 1
\]
Writing $\theta(j_1,j_2,j_3)$ for the unnormalized evaluation of this network,
we see that to achieve this normalization we must divide the projection 
$p \maps j_1 \tensor j_2 \to j_3$ by the square root of $\theta(j_1,j_2,j_3)$.
The formula in Section 9.10 of the book by Kauffman and Lins \cite{KL}
makes it obvious that $\theta(j_1,j_2,j_3)$
is positive (resp.\ negative) when $j_1 + j_2 + j_3$ is
even (resp.\ odd).  This means that when we use normalized trivalent
vertices, a closed trivalent $\SU(2)$ spin network evaluates to a real
number times the product over all vertices $v$ of $i^{j_1 + j_2 +
j_3}$, where $j_1,j_2,j_3$ are the spins labelling the three edges
incident to $v$.  Since the spin label for each edge shows up twice in
this product, it follows that the spin network evaluates to a real
number times $i^{2J}$, where $J$ is sum of the spin labels of all edges.

It therefore suffices to show that whenever the $15j$ symbol is nonzero,
this quantity $J$ is an integer.  To do this, choose any vertex in the
$15j$ symbol spin network and consider its images $v_1, \dots , v_5$
under the action of $\Z_5$ as rotations of the picture.  Each edge is
incident to exactly one of the vertices $v_i$, so $J$ is the sum over $1
\le i \le 5$ of the sum of spins labelling the edges incident to $v_i$.
However, a trivalent vertex can only be nonzero if the sum of spins
incident to it is an integer.  It follows that for the $15j$ symbol to
be nonzero, $J$ must be an integer.  \qed
\vskip 1em

The fact we use here about the $15j$ symbols has a nice generalization:
any trivalent $\SU(2)$ spin network evaluates to real number if its
underlying graph has no loops containing an odd number of edges.  The
reason is that in a nonzero $\SU(2)$ spin network, the edges with
non-integral labels can always be partitioned into disjoint loops.  If
these loops all have an even number of edges, the total of all of the
spins must be an integer, so the above argument shows the spin network
evaluates to a real number.

Using Lemma~\ref{lemma3}, we can easily determine the sign of the usual $10j$ 
symbol:
\begin{lemma} \label{lemma4} 
\[    \TenJ  \quad
 = \quad (-1)^{2(j_1 + \cdots + j_{10})} 
\ModifiedTenJ  \]
It follows that when 
\[    \TenJ \]
is nonzero, it is positive (resp.\ negative) when $j_1 + \cdots + j_{10}$ 
is an integer (resp.\ half-integer).  
\end{lemma}

Proof --- If we go around this graph counterclockwise:
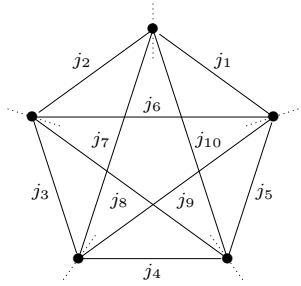
\[ \ModifiedTenJ  \]
each of the ten edges is incoming to exactly one of the modified 
Barrett--Crane intertwiners.  By Lemma \ref{lemma1}, it follows
that 
\[    \TenJ  \quad
 = \quad (-1)^{2(j_1 + \cdots + j_{10})} 
\ModifiedTenJ  \]
The rest follows from Lemma \ref{lemma3}. \qed
\vskip 1em

Finally, we show that these signs cancel when we evaluate a closed
spin foam of the sort that appears in the Barrett--Crane model.  
Here we assume familiarity with the basic definitions concerning
piecewise-linear CW complexes and their homology \cite{Baez,Massey,RS}.

\begin{lemma} \label{lemma5} 
Let $K$ be any 2-dimensional piecewise-linear CW complex, and 
let $c$ be a 2-cycle on $K$ with $\Z_2$ coefficients.  
Let $c(f)$ denote the value of $c$ on any face (i.e., 2-cell)
$f$, and let $n(f)$ denote the number of vertices of the face
$f$ (which equals the number of its edges).  Then $\sum_f n(f) c(f) = 0$.
\end{lemma}

Proof --- Write $f > e$  when the face $f$ has the $e$ as one of its
edges, and let  $(\partial c)(e)$ be the boundary chain of $c$ evaluated
at the edge $e$.  Then 
\[  \sum_f n(f) c(f) = \sum_e \sum_{f > e} c(f) = 
\sum_e  (\partial c)(e) = \sum_e 0 = 0 \] 
\hskip 50 em \qed
\vskip 1em

We now reach our main result.
Here, and in the corollary, we are using the original $10j$ symbols,
so this result is independent of any choices of splittings of vertices.

\begin{theorem} \label{theorem1} 
Let $F$ be a closed spin foam whose underlying 2-dimensional piecewise-linear 
CW complex $K$ is the dual 2-skeleton of simplicial complex formed by
taking a finite set of 4-simplices and gluing them pairwise along all their
tetrahedral faces.   Then $\prod_v A_v \ge 0$, where $v$ runs over the
vertices of $K$ and $A_v$ is the $10j$ symbol associated to $v$.  
\end{theorem}

Proof --- We assume that the product $\prod_v A_v$ is nonzero, and show
that the product of the signs from Lemma \ref{lemma4} is 1.

Since the product is nonzero, each $10j$ symbol is nonzero, and
therefore each Barrett--Crane intertwiner is nonzero.  For each edge in
the spin foam (corresponding to a 3-simplex in the simplicial complex)
there are exactly four faces having it in their boundary
(corresponding to the four triangles in the 3-simplex).  Because the
associated Barrett--Crane intertwiner is nonzero, Lemma \ref{lemma1}
shows that the sum of the spins on these four faces is an integer.  From 
the spin foam $F$ we construct a 2-chain $c$ with values in the
multiplicative group $\{\pm 1\}$  by setting $c(f) = (-1)^{2j_f}$, where
$j_f$ is the spin labelling the face $f$.   The condition above says
that $c$ is a cycle.  Translating Lemma \ref{lemma5} into multiplicative
notation, it follows that $\prod_f c(f)^{n(f)} = 1$. 
By Lemma \ref{lemma4},  the sign of $A_v$ is $\prod_{f>v} (-1)^{2j_f} =
\prod_{f>v} c(f)$, where $f$ runs over all the 2-cells having $v$ as a
vertex.   The sign of $\prod_v A_v$ is therefore  $\prod_v \prod_{f>v}
c(f) = \prod_f c(f)^{n(f)} = 1$. 
\qed
\vskip 1em

In addition to a product of $10j$ symbols, one for every 
vertex, the Barrett--Crane amplitude for a spin foam also involves
a product of factors coming from spin foam edges and faces.  The
precise formulas for these factors are somewhat controversial: 
Perez and Rovelli \cite{PR} have proposed a modification of the
original formulas given by Barrett and Crane \cite{BC}.   However,
in both versions of the Barrett--Crane model, these factors are
{\it nonnegative}.  It follows from the above theorem that 
in both the original Barrett--Crane model and the Perez--Rovelli
version, the amplitude for any closed spin foam is nonnegative.

We can generalize this theorem to open spin foams as follows:

\begin{corollary}\label{cor1}
Let $F \maps \psi\to \psi'$ be an open spin foam whose 
underlying 2-dimensional piecewise-linear 
CW complex is the dual 2-skeleton of a simplicial complex formed by
taking a finite set of 4-simplices and gluing them pairwise along all
their tetrahedral faces except those in which the spin networks 
$\psi$, $\psi'$ lie.  Let $G \maps \psi\to \psi'$ be
another open spin foam of this type, possibly with a different underlying
CW complex.   If the products of $10j$
symbols over the vertices in $F$ and $G$ are both nonzero, 
then these products have the same sign.

\end{corollary}

Proof --- Glue $F$ and $G$ together along $\psi$ and $\psi'$ to get
a closed spin foam $H$.  Since the product of $10j$ symbols for $F$ and
$G$ are both nonzero, the same is true of $H$.  In fact, the result for
$H$ is the product of the results for $F$ and $G$. By
Theorem~\ref{theorem1}, the sign of the evaluation of $H$ is positive,
so the signs of the evaluations of $F$ and $G$ must be the same.
\qed
\vskip 1em

This corollary can be sharpened using an independent argument which
we now sketch.  This sharpened version says that if the evaluation
of $F$ is nonzero, then its sign is $(-1)^{2J}$, where $J$ is the 
sum of the spins labelling the edges of $\psi$ and $\psi'$.
To prove this fact, note that 
any spin foam is a composite of spin
foams with only one vertex, 
since we can slide a cross-section generically from $\psi$ to 
$\psi'$, so that it passes one vertex at a time.
It thus suffices to consider the case where $F$ contains
only one vertex.  
For this, one can examine each of the possible ways in which
$\psi$ can differ from $\psi'$ when there is only a single
vertex in $F$, and note that in each case, the sign of the
$10j$ symbol at that vertex is $(-1)^{2J}$.
For example, a portion of $\psi$ that looks like: 
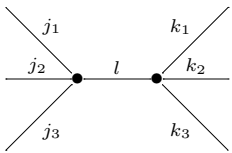
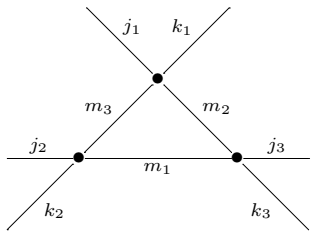
\[
\cxymatrix{
\ar@{-}[dr]^{j_1} & & & \ar@{-}[dl]_{k_1} \\  
\ar@{-}[r]^{j_2} & *{\bullet} \ar@{-}[r]^{l} & *{\bullet} \ar@{-}[r]^{k_2} & \\
\ar@{-}[ur]_{j_3} & & & \ar@{-}[ul]^{k_3}
}
\]
can change to:
\[
\cxymatrix{
& \ar@{-}[dr]^{j_1} & & \ar@{-}[dl]_{k_1} \\
& & *{\bullet} \ar@{-}[dl]_{m_3} \ar@{-}[dr]^{m_2} \\
\ar@{-}[r]^{j_2} & *{\bullet} \ar@{-}[rr]_{m_1} && *{\bullet} \ar@{-}[r]^{j_3} & \\
\ar@{-}[ur]_{k_2} &&&& \ar@{-}[ul]^{k_3}
}
\]
This is the so-called `2--3 Pachner move'.
The sum of the ten spins involved in the $10j$ symbol is
$j_1+j_2+j_3+k_1+k_2+k_3+l+m_1+m_2+m_3$.
Since the evaluation is nonzero, 
$(-1)^{2(j_1+j_2+j_3)}$, $(-1)^{2l}$ and $(-1)^{2(k_1+k_2+k_3)}$ are equal.
Thus the sign of the $10j$ symbol is $(-1)^{2(l+m_{1}+m_{2}+m_{3})}$.
This is equal to $(-1)^{2J}$, where $J$ is the sum of the spins
labelling $\psi$ and $\psi'$, since most of the spins appear twice.

There are other transitions that can occur when the cross section
passes a vertex, such as the 0--5 and 1--4 Pachner moves.
Also, the cross section can pass a cup or a cap in a spin foam edge,
which replaces two vertices with one, or vice versa.
In all of these cases, one can check that the signs work out as
claimed.

\section{Lorentzian Model}

The Barrett--Crane model can be defined for any dimension of spacetime
and any signature \cite{FKP}.  However, our proof of nonnegativity for
spin foam amplitudes in the 4-dimensional Riemannian Barrett--Crane
model relies on two miracles specific to this case: $\Spin(4)$ is a
product of two copies of $\SU(2)$, and its simple representations factor
as $j \tensor j$.  This is what gives the weighted sum of squares in
equation (\ref{sum.of.squares}).  These miracles do not occur in the
4-dimensional Lorentzian model, so we are unable to prove nonnegativity
in this case.  However, we have obtained some numerical evidence for it,
which we present here.

As in the Riemannian case, there is some controversy concerning the
correct edge and face amplitudes in the Lorentzian Barrett--Crane model.
However, in the version for which Crane, Perez and Rovelli \cite{CPR}
have proved convergence of the partition function, these amplitudes are
nonnegative.  This is also true in all the most obvious variants.  Given
this, to prove that the amplitude of a spin foam is nonnegative, it
suffices to show the vertex amplitudes are nonnegative.

The vertex amplitudes in the Lorentzian Barrett--Crane model are
given by the Lorentzian $10j$ symbols \cite{BC2}.  These are obtained
by evaluating $\Spin(3,1)$ spin networks of this form: 
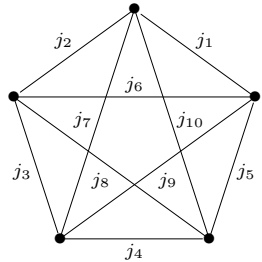
\[   \TenJ \]
Here the edges are labelled by arbitrary nonnegative real numbers
$j_i$, which describe simple
representations of the group $\Spin(3,1) = \SL(2,\C)$.  The vertices are
all labelled with a special intertwiner called the Lorentzian
Barrett--Crane intertwiner.  The simple representation $j$ is the space
of all solutions to the equation $\nabla^2 f= -(j^2+1)f$ on
3-dimensional hyperbolic space that have square-integrable
boundary data on the sphere at infinity.  Since these
representations are infinite-dimensional, a careful procedure must be
used to evaluate the $10j$ spin network.  The end result is
the following integral:
\[ \TenJJ = \frac1 { ({2\pi^2})^4} 
\int_{{\H}^4} \prod_{v,w = 0}^4 
K_{j_{v,w}}(x_v,x_w)\; dx_1 \, dx_2 \, dx_3 \, dx_4 \]
Here $\H$ is 3-dimensional hyperbolic space, 
$dx$ stands for the volume form associated to the
Riemannian metric on $\H$, and $K_j$ is the integral kernel
for the projection onto the simple representation $j$, namely
\[       K_j(x,y) = \frac{\sin jr}{j\sinh r}  \]
where $r$ is the distance betweeen $x,y \in \H$.  
The answer is independent of the free variable $x_{0}$, which
may be held fixed at a convenient point in $\H$.
While this integral has been shown to converge \cite{BB}, no `closed
form' is known, and it is difficult to compute numerically
because it involves 12 variables and an oscillatory integrand.
These difficulties become particularly severe as the $j_i$
become large, since then the integrand oscillates more rapidly.

We have used two methods to compute this integral numerically for
certain choices of $j_i$.  In both methods we start by writing the
integral in hyperbolic spherical coordinates $(r,\theta,\phi)
\in [0,\infty) \times [0,2\pi) \times [0,\pi)$.  Then we do a coordinate
transformation mapping the radial coordinates $r_i$ of the points
$x_1,x_2,x_3,x_4$ to the interval $[0,1)$ via $r_i \mapsto r_i/(1+r_i)$.
Including the Jacobian of this transformation, the result is an 
integral with respect to Lebesgue measure on a product
of four copies of $[0,1) \times [0,2\pi) \times [0,\pi)$.  
Finally, using invariance properties of the integrand, we can
reduce this to a 9-dimensional integral by eliminating the
integrals over $\theta_{1}$, $\phi_{1}$ and $\theta_{2}$.

The first method proceeds by a straightforward Monte Carlo calculation,
uniformly sampling the space over which the integral is being taken, and
averaging the results.  Table 1 lists some results obtained by this
method.  

\vskip 1em

{\vbox{   
\begin{center}   
{\small
\setlength{\extrarowheight}{1.5pt}
\setlength{\tabcolsep}{2.2pt}
\begin{tabular}{|rrrrrrrrrr|D{X}{\ \pm\ }{11}|>{$}r<{$}|}    \hline
$j_1$ & \multicolumn{8}{c}{$\dots$} & $j_{10}$  & \multicolumn{1}{c|}{$10j$ symbol} 
& \multicolumn{1}{c|}{samples} \\ \hline
0 & 0 & 0 & 0 & 0 & 0 & 0 & 0 & 0 & 0   
& 49.34\  X 1.595
& 2.5 \cdot 10^9  \\  \hline

1 & 1 & 1 & 1 & 1 & 1 & 1 & 1 & 1 & 1   
& 9.420 \cdot 10^{-3}\  X 3.964 \cdot 10^{-5}
& 1.0 \cdot 10^9  \\   \hline

2 & 2 & 2 & 2 & 2 & 2 & 2 & 2 & 2 & 2   
& 9.138 \cdot 10^{-6}\  X 1.268 \cdot 10^{-7}
& 1.0 \cdot 10^8   \\   \hline

9 & 9 & 9 & 9 & 9 & 9 & 9 & 9 & 9 & 9   
& 3.937 \cdot 10^{-13} X 4.167 \cdot 10^{-14}
& 1.6 \cdot 10^9  \\   \hline

19 & 1 & 1 & 1 & 1 & 1 & 1 & 1 & 1 & 1  
& 2.017 \cdot 10^{-6} \  X 1.498 \cdot 10^{-6}
& 9.0 \cdot 10^9   \\   \hline

1 & 2 & 3 & 4 & 5 & 1 & 2 & 3 & 4 & 5   
& 1.368 \cdot 10^{-8}\  X 1.814 \cdot 10^{-9}
& 7.2 \cdot 10^9 \\   \hline

1.1 & 1.3 & 0.4 & 2.1 & 1.6 & 1.5 & 0.9 & 1.2 & 1.9 & 0.8 
& 4.884 \cdot 10^{-4}\  X 1.046 \cdot 10^{-5}
& 8.0 \cdot 10^7  \\   \hline

1.4 & 15 & 0.1 & 3 & 2 & 1 & 1 & 1 & 0.1 & 8 
& 1.100 \cdot 10^{-7}\  X 1.670 \cdot 10^{-7}
& 5.2 \cdot 10^8  \\   \hline
\end{tabular}} 
\end{center}   
\vskip 1em 
\centerline{Table 1: Lorentzian $10j$ symbols --- Monte Carlo results}  
}}   
\vskip 1em   

The second method uses the `Vegas' algorithm.  This is an adaptive
algorithm that combines importance sampling and stratified sampling.
The rather complicated details are explained in Section 7.8 of the book
{\sl Numerical Recipes} \cite{NR}.  Table 2 lists some results obtained
by this method.  Each entry in the table represents a weighted average
of independent runs of the Vegas algorithm.

\vskip 1em   
{\vbox{   
\begin{center}   
{\small
\setlength{\extrarowheight}{1.5pt}
\setlength{\tabcolsep}{2.2pt}
\begin{tabular}{|rrrrrrrrrr|D{X}{\ \pm\ }{11}|>{$}l<{$}|}    \hline
$j_1$ & \multicolumn{8}{c}{$\dots$} & $j_{10}$  & \multicolumn{1}{c|}{$10j$ symbol} 
& \multicolumn{1}{c|}{samples} \\ \hline
0 & 0 & 0 & 0 & 0 & 0 & 0 & 0 & 0 & 0   
& 53.42 X .0689
& 5.0 \cdot 10^8  \\  \hline

1 & 1 & 1 & 1 & 1 & 1 & 1 & 1 & 1 & 1   
& 9.526 \cdot 10^{-3}\  X 1.367 \cdot 10^{-6}
&  7.5 \cdot 10^8   \\   \hline

2 & 2 & 2 & 2 & 2 & 2 & 2 & 2 & 2 & 2   
& 9.095 \cdot 10^{-6}\  X 8.153 \cdot 10^{-9}
& 5.0 \cdot 10^8  \\   \hline

9 & 9 & 9 & 9 & 9 & 9 & 9 & 9 & 9 & 9   
& 3.626 \cdot 10^{-13} X 1.226 \cdot 10^{-15}
& 1.3 \cdot 10^9  \\   \hline

19 & 1 & 1 & 1 & 1 & 1 & 1 & 1 & 1 & 1  
& 5.602 \cdot 10^{-8} \  X 6.934 \cdot 10^{-8}
& 6.5 \cdot 10^9  \\   \hline

1 & 2 & 3 & 4 & 5 & 1 & 2 & 3 & 4 & 5   
& 1.162 \cdot 10^{-8}\  X 6.057 \cdot 10^{-11}
& 1.7 \cdot 10^{10}  \\   \hline

1.1 & 1.3 & 0.4 & 2.1 & 1.6 & 1.5 & 0.9 & 1.2 & 1.9 & 0.8 
& 5.029 \cdot 10^{-4}\  X 9.258 \cdot 10^{-7}
& 5.0 \cdot 10^7  \\   \hline

1.4 & 15 & 0.1 & 3 & 2 & 1 & 1 & 1 & 0.1 & 8 & 
3.068 \cdot 10^{-9}\  X 6.860 \cdot 10^{-9}
& 5.5 \cdot 10^8  \\   \hline

\end{tabular}} 
\end{center}   
\vskip 1em 
\centerline{Table 2: Lorentzian $10j$ symbols --- Vegas results}  
}}   
\vskip 1em   

In both these tables, the error estimates are calculated as
\begin{equation}
 \Delta f = V \sqrt{\frac{\langle f^2\rangle - \langle f\rangle^2}{N}}, 
\label{error}
\end{equation}
where $\langle f \rangle$ denotes the mean of the samples of the
function $f$, $N$ is the number of samples, and $V$ is the volume of the
region of integration.  In the case of the Vegas algorithm the samples
are not uniformly chosen, so the mean is a weighted one, as explained
in {\sl Numerical Recipes}.

The reader can easily note a few things from these tables.  First,
the computed $10j$ symbols are always nonnegative.  In two cases,
namely 
\[    (j_1, \dots, j_{10}) = 
(19, 1, 1, 1, 1, 1, 1, 1, 1, 1)  \]
and
\[    (j_1, \dots, j_{10}) = 
(1.4, 15, 0.1, 3, 2, 1, 1, 1, 0.1, 8)  \]
the computed $10j$ symbols are smaller than their error bars,
consistent with them being zero or even negative.  However, 
this is understandable, since there is a good reason to expect 
the $10j$ symbols to be very small in these cases.  The reason 
is that the Barrett-Crane intertwiner 
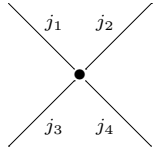
\[     \FourX{j_1}{j_2}{j_3}{j_4}  \]
becomes exponentially damped when the spins $j_1,\dots,j_4$ leave the
region defined by the `tetrahedron inequalities':
\[
\begin{array}{ccc}
j_1 &\le& j_2 + j_3 + j_4, \\
j_2 &\le& j_1 + j_3 + j_4, \\
j_3 &\le& j_1 + j_2 + j_4, \\
j_4 &\le& j_1 + j_2 + j_3. 
\end{array}
\]
This is most easily seen from the explicit formula in \cite{BC2}
for the value of the $\Spin(4)$ spin network
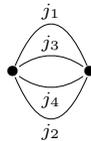
\[
\fourtheta{j_1}{j_2}{j_3}{j_4}
\]
which can be thought of as the norm squared of the Barrett--Crane
intertwiner. 

Second, the Vegas results generally have smaller error bars than the 
corresponding Monte Carlo results, and usually lie within the error
bars of the Monte Carlo results.  This suggests (but does not
rigorously prove) that the Vegas algorithm gives more accurate
values for the $10j$ symbols.  

Finally, the large sample sizes required to obtain even tolerable
accuracy show the difficulty of these numerical computations.  It will
be important to develop new theoretical methods to prove or disprove
the conjectured nonnegativity of these Lorentzian $10j$ symbols,  and
also new algorithms to more efficiently compute them.   

\section{Conclusions}

Given the results here together with the efficient algorithm for
computing the Riemannian $10j$ symbols described in a companion paper
\cite{CE}, it is now practical to study the Riemannian Barrett--Crane
model using the Metropolis algorithm.  Some results along these lines
will appear in another paper \cite{BaezChristensen}.  For the Lorentzian
model we still need better formulas for the $10j$ symbols, both for
proving nonnegativity and for numerical computations.

The physical significance of Corollary 1 is still somewhat mysterious.
Multiplying the usual spin network states by
suitable signs, it means we can find 
orthonormal bases of the kinematical Hilbert
spaces of any triangulated compact 3-manifolds $S$ and $S'$ such that for
any triangulated cobordism $M\maps S \to S'$, the matrix elements of the
operator $Z(M) \maps Z(S) \to Z(S')$ are always a sum of nonnegative
terms, one for each spin foam $F \maps \psi \to \psi'$ in the dual
2-skeleton of $M$.  At first it was suggested that this means there is
no destructive interference between spin foams in the Riemannian
Barrett-Crane model.  Subsequent discussions have clarified this issue
somewhat.  Indeed, if a unitary operator has all nonnegative matrix
elements in some basis it must be rather trivial, simply acting as a
permutation of the basis.  However, thanks to the `problem of time'
in quantum gravity, it is more natural to use a sum over spin foams
to compute, not a time evolution operator, but the
projection from the space spanned by spin networks onto the
physical Hilbert space of quantum gravity \cite{Arnsdorf,Rovelli}.  
There is no reason for this to be unitary; it should be something
more like a positive operator.  There is thus no contradiction if it
has nonnegative matrix elements.  Indeed, this also occurs in other
theories, such as the quantum mechanics of a relativistic free particle,
and also quantum gravity in $2+1$ dimensions.  However, we still lack
a general principle explaining why all spin foams $F \maps \psi \to \psi'$
should have real amplitudes of the same sign.

\section*{Acknowledgements}

We would like to thank John Barrett, Greg Egan and Tom Halford for
conversations and code related to this project.  We also thank Lee
Smolin for drawing attention to the need for numerical methods suited to
real-time path integrals, Rafael Sorkin for raising the issue of a
possible lack of destructive interference in models with positive spin
foam amplitudes, and Abhay Ashtekar, Alejandro Perez, and Thomas
Thiemann for clarifying this issue.

\end{document}